\documentclass[twocolumn,linenumbers]{aastex631}

\defcitealias{Adegoke2019}{OA2019}
\defcitealias{Buisson2018}{B18}
\defcitealias{Robertson2015}{R15}
\begin{document}
\nolinenumbers

\title{A Search for X-ray/UV Correlation in the Reflection-Dominated Seyfert 1 Galaxy Mrk 1044 }% \footnote{Released on March, 1st, 2021}}

\author{Samuzal Barua}
\affiliation{Department of Physics, Gauhati University, Jalukbari, Guwahati 781014, Assam, India}

\author{Oluwashina K. Adegoke}
\affiliation{Cahill Center for Astronomy and Astrophysics, California Institute of Technology, Pasadena, CA 91125, USA}

%\collaboration{20}{(AAS Journals Data Editors)}

\author{Ranjeev Misra}
\affiliation{Inter-University Centre for Astronomy and Astrophysics (IUCAA), PB No. 4, Ganeshkhind, Pune 411007, India}

\author{Pramod Pawar}
\affiliation{Inter-University Centre for Astronomy and Astrophysics (IUCAA), PB No. 4, Ganeshkhind, Pune 411007, India}

\author{V. Jithesh}
\affiliation{Department of Physics and Electronics, CHRIST (Deemed to be University), Hosur Main Road, Bengaluru 560029, India}

\author{Biman J. Medhi}
\affiliation{Department of Physics, Gauhati University, Jalukbari, Guwahati 781014, Assam, India}



\begin{abstract}
\nolinenumbers
Correlated variability between coronal X-rays and disc optical/UV photons provides a very useful diagnostic of the interplay between the different regions around an active galactic nucleus (AGN) and how they interact. AGN that reveal strong X-ray reflection in their spectra should normally exhibit optical/UV to X-ray correlation consistent with reprocessing -- where the optical/UV emission lag behind the X-rays. While such correlated delay has been seen in some sources, it has been absent in others. \rm{Mrk~1044} is one such source that has been known to reveal strong X-ray reflection in its spectra. In our analysis of three long \textit{XMM-Newton} and several \textit{Swift} observations of the source, we found no strong evidence for correlation between its UV and X-ray lightcurves both on short and long time scales. Among other plausible causes for the non-detection, we posit that higher X-ray variability than UV and strong general relativistic effects close to the black hole may also be responsible. We also present results from the spectral analysis based on \textit{XMM-Newton} and \textit{NuSTAR} observations, which show the strong soft X-ray excess and iron K$\alpha$ line in the 0.3--50 keV spectrum that can be described by relativistic reflection.
%{\bf Results from the spectroscopic analysis is also presented, denoting that 0.3--50 keV spectrum can be described by relativistic reflection.}
\end{abstract}



%%% \latex\ \footnote{\url{http://www.latex-project.org/}} 

\section{Introduction} \label{sec:intro}

Emission variability in active galactic nuclei (AGN) has been observed over the entire electromagnetic spectrum -- from radio to gamma rays -- and over an extensive range of timescales \citep{McHardy1999,  Uttley2004, Breedt2010}.

AGN host a supermassive black hole (SMBH) at their centres. Accretion of matter onto the black hole powers the AGN, turning gravitational energy into kinetic and viscous internal energy. As a result, the accretion disc emits thermal radiation, mostly seen in optical/UV energies \citep{Koratkar1999}. 
The X-ray emission from AGN has been interpreted as a consequence of inverse-Compton scattering of seed optical/UV photons from the accretion disc interacting with energetic electrons in a ``corona'' \citep{Haardt1993, Merloni2003, Fabian2015, Wilkins2015, 2017MNRAS.466.3951A, Samuzal2020}.  

It has been shown over the past few decades that emission in the different energy bands around AGN tend to correlate in their variability for many sources. Of particular interest is the correlation in variability between emission in the optical/UV -- produced in the geometrically thin, optically thick disc \citep{1973A&A....24..337S, 1973blho.conf..343N} -- and those in the X-rays. 
Published results from such emission variability studies have shown varying trends, which sometimes are also a function of the timescale considered. Some sources show correlated variability such that the X-ray photons lead the optical/UV photons and this is normally interpreted as indicative of reprocessing -- where the X-rays have been reprocessed into the observed optical/UV photons after reflection in the disc \citep[see e.g.,][]{2003ApJ...584L..53U, 2009MNRAS.397.2004A, Cameron2012, McHardy2014, Edelson2015, Pal2017, Buisson2017, Lobban2018}. Systems that exhibit this sense of variability are additionally expected to show signatures consistent with a reflection-dominated spectrum, e.g. broad iron $K_{\alpha}$ emission features seen at $\sim6.4\,\mathrm{keV}$ \citep{2004MNRAS.353.1071F}, since strong X-ray reflection components naturally imply that the X-rays strongly illuminates the inner disc. 

For a handful of sources, the optical/UV emission lead the X-rays in their variability and this has been argued to result from the inverse Compton scattering of disc optical/UV photons into X-rays \citep{2005A&A...430..435A, Adegoke2019} or the inward propagation of fluctuations through the accretion disc \citep{2008ApJ...677..880M, 2013MNRAS.433.1709G}. Other sources either reveal a complex X-ray/UV correlated behaviour \citep[e.g.,][]{Pawar2017, Pal2018, Kumari2021} or a correlation consistent with zero lag \citep[e.g.,][]{Breedt2009}.

 A few other sources still, show no measurable correlation in their optical/UV and X-ray variabilities. For example, in the multi-epoch \textit{XMM-Newton} observations of the narrow-line Seyfert 1 (NLS1) AGN \rm{1H~0707-495}, \citet[][hereafter \citetalias{Robertson2015}]{Robertson2015} found no correlation between the UV and the X-ray emission variability. Also, \citet[][hereafter \citetalias{Buisson2018}]{Buisson2018} detected no physically plausible correlation from the \textit{Swift} and \textit{XMM-Newton} observations of the highly variable NLS1 AGN \rm{IRAS~13224-3809}. This is despite the fact that both systems have been known to show strong signatures of reflection in their X-ray spectra. 
 
\rm{Mrk~1044} is a nearby highly accreting NLS1 AGN at a redshift of $z\sim0.016$ that has been known for its peculiar variability properties having been observed multiple times in different energy bands \citep[see e.g.,][]{Wang2001, Dewangan2007, Du2015, Mallick2018, Gliozzi2020}. Using a short \textit{XMM-Newton} observation of \rm{Mrk~1044}, \citet{Dewangan2007} found no evidence for a lag between the $0.2-0.3\,\mathrm{keV}$ and the $5-10\,\mathrm{keV}$. X-ray bands on the basis of which they ruled out reprocessing as the plausible origin of its prominent soft X-ray excess. 
From the broadband 2013 \textit{XMM-Newton} and 2016 \textit{NuSTAR} observations of the source,  \citet{Mallick2018} found prominent soft and hard X-ray excesses as well as iron $K_{\alpha}$ line well described by the relativistic reflection model. The observations revealed that the X-ray continuum of \rm{Mrk~1044} can adequately be described by the relativistic reflection model similar to those of \rm{1H~0707-495} and \rm{IRAS~13224-3809}. The black hole mass of Mrk 1044 has been measured through reverberation mapping campaign to be $\sim3\times10^{6}\,\mathrm{M_{\odot}}$ \citep{Wang2001, Du2015}. Couple of studies quantified the accretion rate ($L_{bol}/L_{Edd}$) that is ranging over 1.2 \citep{Husemann2022} to 16 \citep{Du2015}. Applying the bolometric correction term ($k=L_{bol}/L_{2-10~keV}$) for 2--10 keV X-ray luminosity, the accretion rate has also been found to $\sim0.3$ \citep{Laha2018}.

%%%%%%%%%%%%%%%%%%%%%%%%%%%%%%%%%%%%%%%%%%%%%%%%%%%%%%%%%%%%%%%%%%%%%%%%%%%%%%%%%%%%%%
\begin{figure}
	\hspace{-0.3cm}
	\includegraphics[width=8.8cm, height=6.5cm]{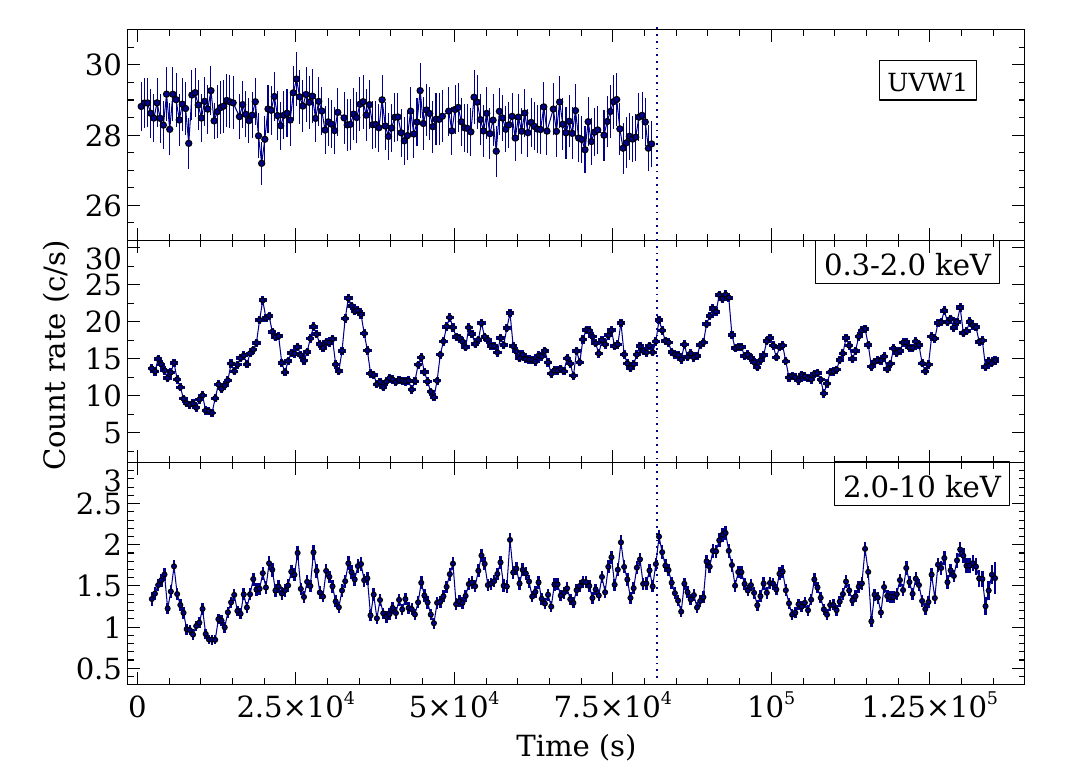}
	\vspace{0.05cm}
    \includegraphics[width=8.2cm, height=6.2cm]{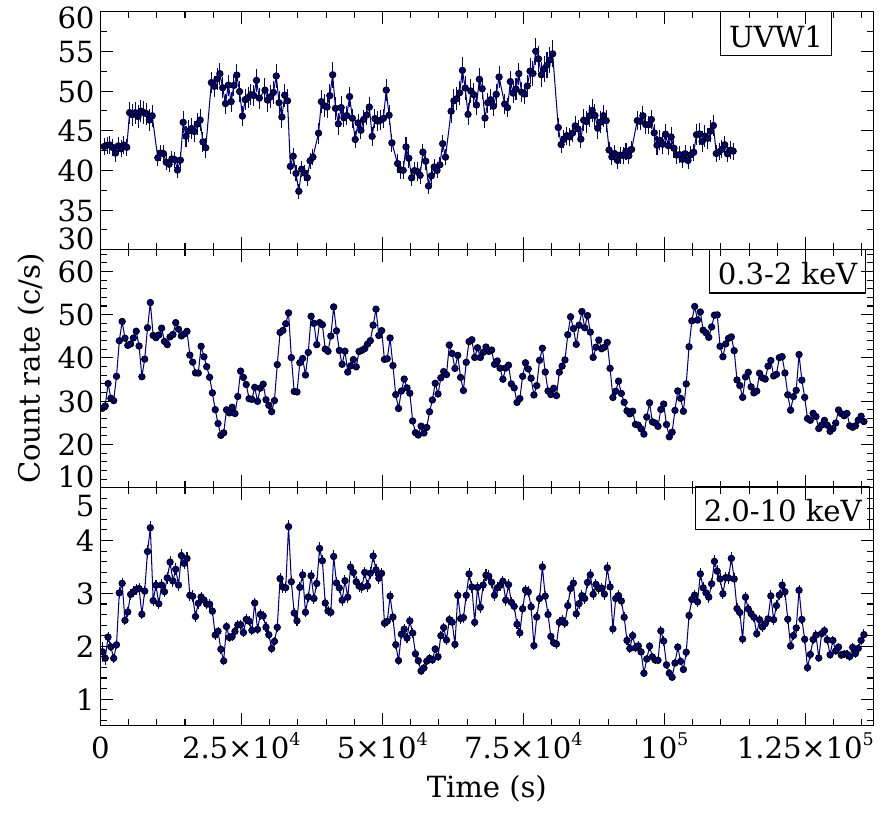}
	\vspace{0.05cm}
	\includegraphics[width=8.2cm, height=6.5cm]{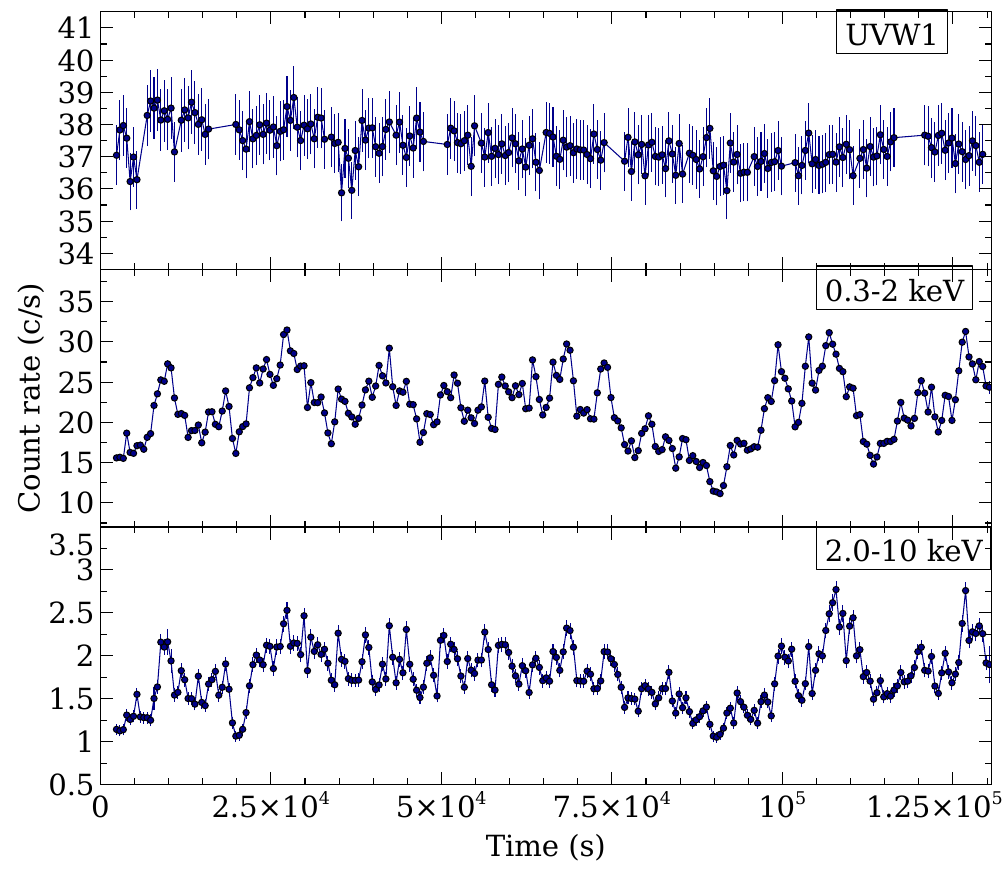}
	\caption{\textit{XMM-Newton} lightcurves of Mrk 1044 from XMM13 ({\it top}), XMM18 ({\it middle}) and XMM19 ({\it bottom}) observations. All lightcurves are obtained using $500\,\mathrm{s}$ time bin. Panels from top to bottom, each of the plots depict lightcurves for UVW1, soft X-ray (0.3--2 keV) and the hard X-ray (2--10 keV) bands, respectively.}
	\label{fig:fig1}
\end{figure}
\begin{table*}
	\centering
	\caption{{\it XMM-Newton}, {\it Swift}, {\it NuSTAR} observations and corresponding count rates (c/s) and fractional variability amplitudes ($F_{var}$) estimated for \rm{Mrk 1044}. The estimated values are represented for UV, soft X-ray ($0.3-2\,\mathrm{keV}$), hard X-ray ($2-10\,\mathrm{keV}$) and the total X-ray ($0.3-10\,\mathrm{keV}$) bands. The errors associated with each parameter are calculated at the $3\sigma$ confidence level.}
	\label{tab:table1}
	\begin{tabular}{lccccccc} % four columns, alignment for each
		\hline
		Observation ID & Name & X-ray/UV exposure (ks) & Estimations &UV & SX & HX & Total X-ray\\
		\hline
		0695290101&XMM13&133/80 & count rate (c/s)& 28.46$\pm$0.06 & 15.75$\pm$0.02 & 1.5$\pm$0.01 & 17.22$\pm$0.02\\
		\vspace{0cm}\\
		& & &$F_{var}(\%) $&2.42$\pm$0.27 & 19.8$\pm$0.09 & 15.39$\pm$0.07 & 19.27$\pm$0.08\\
		\vspace{0.0cm}\\
        0824080301& XMM18 & 140/132 & count rate (c/s)& 45.25$\pm$0.24 & 37.10$\pm$0.02 & 2.67$\pm$0.06 & 39.13$\pm$0.50\\
		\vspace{0cm}\\
		&&&$F_{var}(\%) $&8.00$\pm$0.02 & 20.40$\pm$0.10 & 20.93$\pm$0.09 & 21.31$\pm$0.05\\
        \vspace{0cm}\\
		0841820201& XMM19&130/130 & count rate (c/s)& 37.33$\pm$0.06 & 21.76$\pm$0.02 & 1.77$\pm$0.01 & 23.53$\pm$0.02\\
		\vspace{0cm}\\
		&&&$F_{var}(\%) $&1.8$\pm$0.35 & 19.22$\pm$0.08 & 18.40$\pm$0.01 & 19.00$\pm$0.01\\
        \vspace{0cm}\\
        00035760007-31& Swift &24.6/24.3 & count rate (c/s)& 60$\pm$ 2.01 & 1.66$\pm$0.15 & 0.45$\pm$0.03 &  2.11$\pm$0.18\\
		\vspace{0cm}\\
		&&&$F_{var}(\%) $&17.3$\pm$0.005 & 46$\pm$0.01 & 36$\pm$0.02 & 43.3$\pm$0.01\\
        \vspace{0cm}\\
        60401005002& NuSTAR & 267  & count rate (c/s)& -- & -- & -- &  0.7$\pm$0.0.01\\
        \vspace{0cm}\\
    %    00035760007-31& Swift & 
		\hline
		\hline
	\end{tabular}
\end{table*}
\begin{figure*}
\centering
	%	\begin{minipage}{.45\textwidth}
     \hspace{-0.5cm}
	\includegraphics[width=6cm, height=6cm]{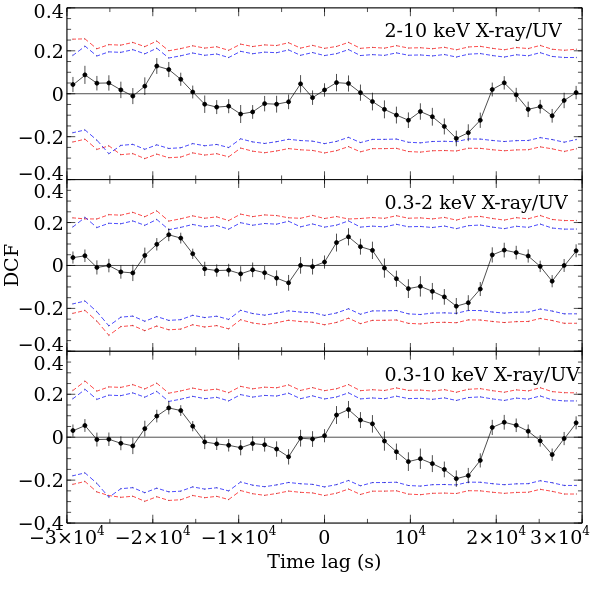}
    \hspace{0.01cm}
    \includegraphics[width=6cm, height=6cm]{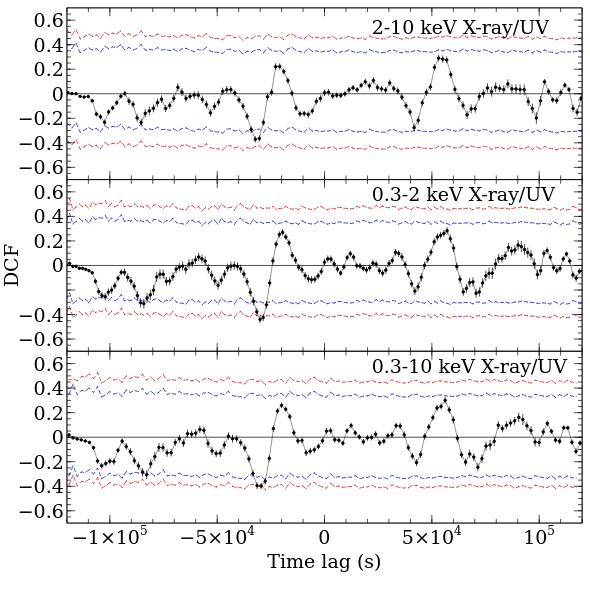}
	\hspace{0.01cm}
	%	\begin{minipage}{.45\textwidth}
	\includegraphics[width=6cm, height=6cm]{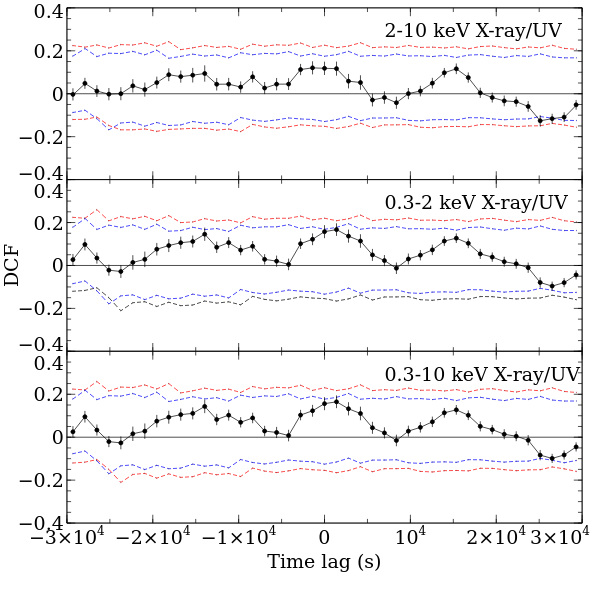}
	\caption{Plots from the implementation of Discrete Correlation Function (DCF) between X-ray and UV lightcurves. From left to right the plots are produced from XMM13, XMM18 and XMM19 data, respectively. The top panels of each plot represent DCF of HX/UV lightcurves, bottom left panels for SX/UV and bottom right panels represent DCF for full $0.3-10\,\mathrm{keV}$ X-ray/UV lightcurves. The blue and red dashed lines represent confidence intervals at 95\% and 99\%, respectively}.  
	\label{fig:fig2}
	%   \end{minipage}
\end{figure*}
%%
%%%%%%%%%%%%%%%%%%%%%%%%%%%%%%%%%%%%%%%%%%%%%%%%%%%%%%%%%%%%%%%%%%%%%%%%%%%%%%%%%%%%%

Here, we investigate short-term and long-term correlated X-ray/UV variability in {Mrk~1044} using three long {\it XMM-Newton} and numerous {\it Swift} observations. We further explore X-ray reflection dependency of the source via spectroscopic analysis with one of the {\it XMM-Newton} and longest {\it NuSTAR} observations.

This paper is organised as follows: In Section 2, we discuss the observation and data reduction procedure. Section 3 includes a detailed discussion on the correlation analysis, methods and the results. Section 4 presents the X-ray spectral analysis. In Section 5, we provide interpretation for our results and then conclude with section 6.  

\section{Observations and Data Reduction}

\subsection{{\it XMM-Newton}}
 Mrk 1044 has been observed several times by \textit{XMM-Newon} \citep{Jansen2001}. The $\sim130-140\,\mathrm{ks}$ observations of the source carried out in 2013, 2018 and 2019 (Observation IDs: 0695290101, 0824080301 and 0841820201 respectively) are particularly useful in studying X-ray/UV variability and their plausible correlations. This is because the observations made use of both the European Photon Imaging Cameras (EPIC) \citep{Kendziorra2001} in the X-ray band and the Optical Monitor (OM) \citep{Mason2001} observing in the optical/UV bands -- with the OM observations carried out in the \textit{image+fast} mode. Therefore, we have used these three observations in our analysis. Details of these observations are presented in Table~\ref{tab:table1} and are named XMM13 (0695290101), XMMM18 (0824080301) and XMM19 (0841820201). During all three observations, the EPIC-pn was operated in the small window mode using thin filter for XMM13 and medium filters for XMM18 and XMM19. For XMM13, the OM observation was carried out using the U, B, V, UVW1, UVW2, and UVM2 filters while only the UVW1 filter was used during XMM18 and XMM19. Here, we have used the UVW1 OM observations (with effective wavelength $\lambda_{eff}=2910\textup{\AA}$) having total exposure times of $\sim80\,\mathrm{ks}$ from XMM13, $\sim132\,\mathrm{ks}$ for XMM18 and $\sim130\,\mathrm{ks}$ from XMM19 alongside the EPIC-pn X-ray observations of $\sim130\,\mathrm{ks}$ duration for XMM13, $140\,\mathrm{ks}$ for XMM18 and $\sim130\,\mathrm{ks}$ for XMM19. The UVW1 observations comprise a collection of 20 short exposures for XMM13, 29 for XMM18 and 25 exposures for XMM19. Other \textit{XMM-Newton} observations of \rm{Mrk~1044} were not usable either due to minimal exposure as in the 2002 observation, unphysical trend in the UV lightcurves or issue in X-ray detection as evident in the other 2018 observations. 

\begin{figure*}
\centering
        \hspace{-0.5cm}
		\includegraphics[width=6cm, height=6cm]{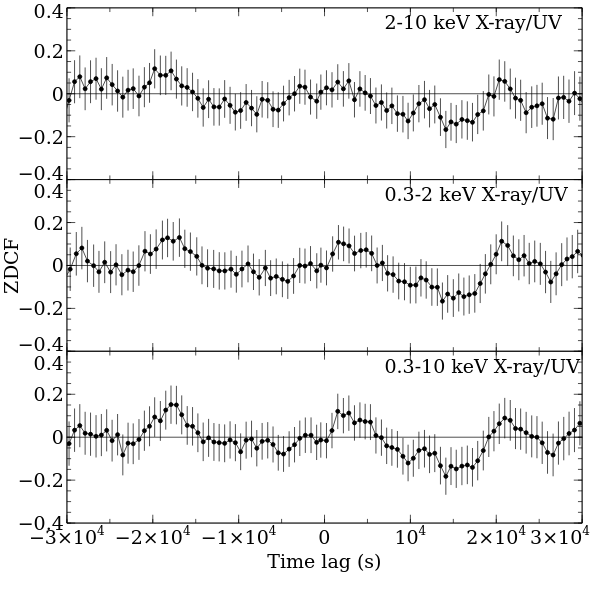}
	      \hspace{0.1cm}
        \includegraphics[width=6cm, height=6cm]{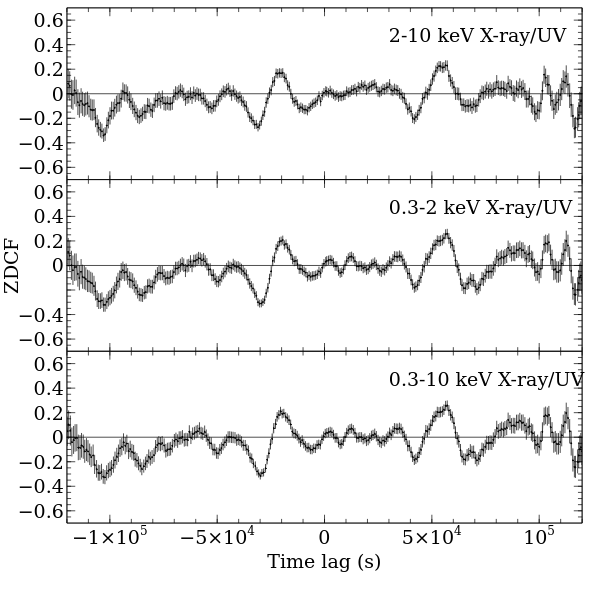}
		\includegraphics[width=6cm, height=6cm]{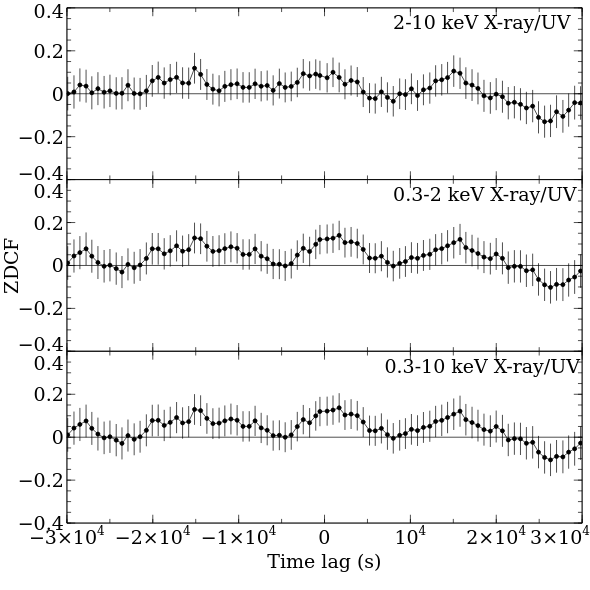}
		\caption{Plots of z - transformed discrete correlation functions (ZDCF) obtained for XMM13 (left), XMM18 (middle) and XMM19 (right) observations. From top to bottom panels of each figure, ZDCF plots are depicted for hard X-ray/UV, soft X-ray/UV and entire X-ray/UV bands.  
		}
		\label{fig:fig3}
\end{figure*}

The \textit{XMM-Newton} data files were processed with the Science Analysis System (\texttt{sas v.17.0.0}) using updated Current Calibration Files (CCF). The event file lists were extracted using the {\sc evselect} task. Afterwards, the data sets were screened and filtered for any high background flares, following which a good time interval (gti) file was created. Using this gti file with the event files, a clean event file was produced again with the {\sc evselect} tool. With this clean event file, EPIC-pn image files were created from which circular source regions of 50 arcseconds and nearby background circular regions of 60 arcseconds were extracted from the same CCDs. Adopting these regions from the image files, spectra and lightcurve products were generated for both the sources and their respective backgrounds. We used a time bin size of $500~$s for creating the lightcurves. Background-subtracted lightcurves were then created using the {\sc epiclccorr} task in three different X-ray energy bands, $0.3-2\,\mathrm{keV}$ (soft band), $2-10\,\mathrm{keV}$ (hard band) and $0.3-10\,\mathrm{keV}$ (full band) respectively. %These are designated as soft X-ray (0.3-2 keV), hard X-ray (2-10 keV) and full X-ray bands.
Fig.~\ref{fig:fig1} shows the representative lightcurevs for soft and hard X-ray bands. By applying the sas tasks {\sc rmfgen} and {\sc arfgen} respectively, we constructed rmf and arf files. The spectra were then grouped using the {\sc specgroup} command to have at least 20 counts per spectral bin.

The SAS meta-task {\sc omfchain} was used to extract the UVW1 event list and to generate the lightcurves using $500\,\mathrm{s}$ binning. For the respective observations, all the UVW1 lightcurve files from the individual exposures were combined to obtain net UVW1 lightcurves. The UVW1 lightcurves are displayed in the top panels of Fig.~\ref{fig:fig1}.  
%%%%%%%%%%%%%%%%%%%%%%%%%%%%%%%%%%%%%%%%%
%%
\begin{figure}
\centering
		\includegraphics[width=8.5cm, height=6.5cm]{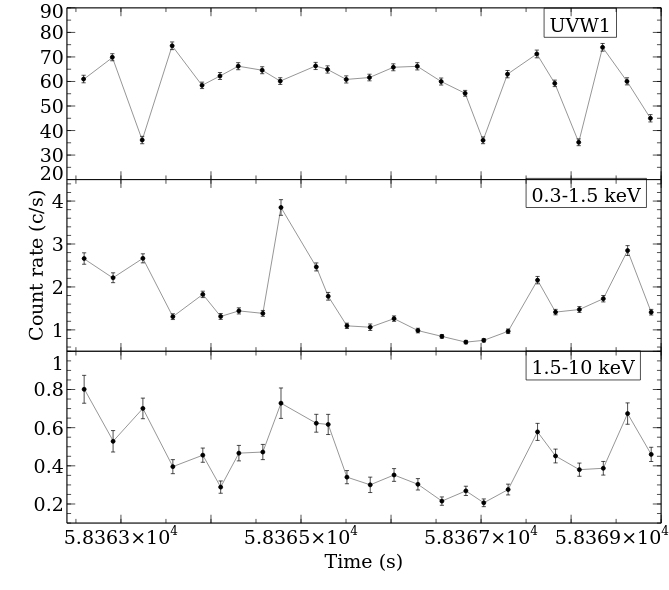}
		\caption{{\it Swift} lightcurves of Mrk 1044 from 2018 observations. Top, middle and bottom panels show lightcurevs for UVW1 band, soft X-ray (0.3--1.5 keV) band and hard X-ray (1.5--10 keV) band, respectively. }
		\label{fig:swift_lcurve}
\end{figure}
\subsection{\it Swift}
{\it Swift} \citep{Gehrels2004} has multiple optical/UV and X-ray observations of Mrk 1044 spanning several years. In this work, we have used data from the monitoring campaign of 2018 when the source was observed for several days in August. Table~\ref{tab:table1} includes details about the observations used. For UVOT \citep{Roming2005}, the lightcurves were extracted from the image files generated as part of the level II data. We employed {\sc UVOTSOURCE} to perform photometry on the image products using the UVW1 filter which has about the same effective wavelength as the OM-UVW1 filter in {\it XMM-Newton}. For the extraction of the lightcurve, a circular source region of radius 5 arcsec is chosen and a larger source free region of radius 15 arcsec is selected for the background. Following the procedure outlined in \citet{2015ApJ...806..129E}, the UVOT data were filtered by eliminating the exposures in which the source region encounters certain areas of the detector that result in low counts. Finally, all the short exposures were summed to obtain the net lightcurve. UVW1 lightcurve from {\it Swift} is displayed in the top panel of Fig.~\ref{fig:swift_lcurve}. 
%{\bf Using the same source and background regions we also created PHA files for UVOT by employing the {\sc uvot2pha} command.}

{\it Swift} XRT \citep{Burrows2005} lightcurves were generated in three different energy ranges; $0.3-1.5 \,\mathrm{keV}$ (soft X-ray), 1.5--10 keV (hard X-ray) and 0.3--10 keV (full X-ray). We show the soft/hard X-ray lightcurves in Fig.~\ref{fig:swift_lcurve}. The X-ray data were obtained in the photon counting (PC) mode. We used the publicly available {\it Swift-XRT products}\footnote{\url{https://www.swift.ac.uk/user_objects/}} \citep{Evans2007, Evans2009} to obtain these lightcurves. The selection criterion imposed for the extraction regions is that the source is of 1.2 arcmin, and an annulus background region of radii from 2.3 to 7 arcmin.  

\subsection{\it NuSTAR}
Mrk 1044 has been observed by {\it NuSTAR} a number of times. We use the longest one obtained from 2018 (ObsID: 60401005002; see Table~\ref{tab:table1}). This observation was carried out for a duration of $\sim267\,\mathrm{ks}$. The data were reduced using the standard {\it NuSTAR} data analysis software \textsc{nustardas v1.8.0} and \textsc{caldb v20171002}. While the \texttt{nupipeline} command was employed to generate the cleaned event files and images, the \texttt{nuproducts} meta-task was used to generate the final products including the lightcurves, source and background spectra as well as the associated response files. The source and background regions used with \texttt{nuproducts} were of 90 arcsec and 100 arcsec radii respectively. The spectra were then grouped using \textsc{grppha} to have a minimum of 50 counts per spectral bin. 

%We reduced the {\it NuSTAR} data for any high flare by using {\sc NUPIPILINE} task and obtained cleaned and calibrated event files. The reduction process required the use of the standard {\it NuSTAR} data analysis software {\sc NUSTARDAS}  (v1.8.0) and CALDB with its version 20171002. Using the {\sc NUPRODUCTS} script, we extracted the final data products, such as, the spectra and the lightcurves from the cleaned events. To have these products, a source extraction region of radius 90 arcsec and a slightly larger background region of 100 arcsec radius were selected for both the FPMA and FPMB (Focal Plane Module A and Focal Plane Module B) instruments.  Finally, the spectra were grouped with the use of {\sc grppha} tool taking 50 counts per spectral bin.
%%%%
\begin{figure*}
\centering
		\includegraphics[width=6.5cm, height=6cm]{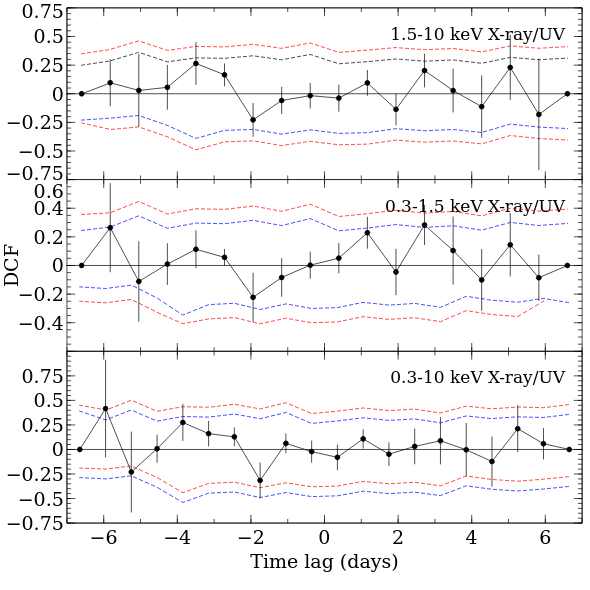}
	\hspace{0.7cm}
		\includegraphics[width=6.5cm, height=6cm]{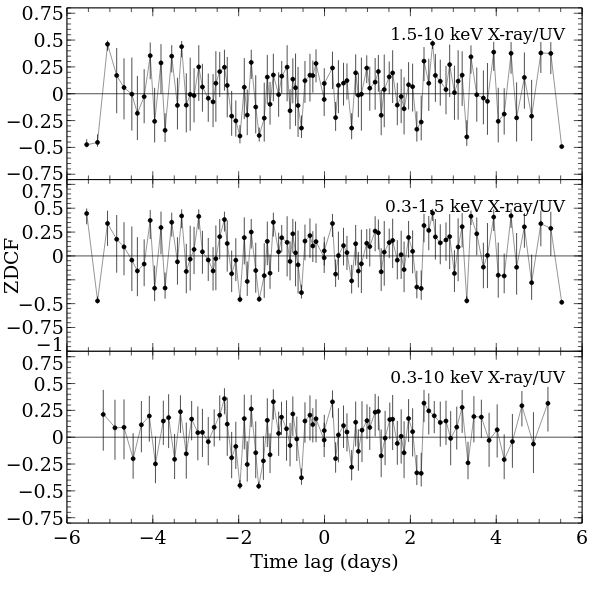}
		\caption{DCF (left) and ZDCF (right) of the {\it Swift} XRT lightcurves against UVOT. In the plots, panels from top to bottom represent correlation functions for hard X-ray, soft X-ray and full X-ray bands versus UV band.}
		\label{fig:fig4}
\end{figure*}
\section{Lightcurves, Correlation Analysis and Results}
\subsection{Short time scale correlation}
%It is evident from the \textit{XMM-Newton} lightcurves shown in  Figure~\ref{fig:fig1} that \rm{Mrk~1044} is highly variable in its emission from the UV to X-ray energy bands. 
From visual inspection on the \textit{XMM-Newton} lightcurves shown in  Figure~\ref{fig:fig1} it is clear that Mrk 1044 exhibits significant X-ray variability on short timescales for all three Epochs of observation. On the other hand, while the UV lightcurve from XMM18 also show significant variability, the UV lightcurves of XMM13 and XMM19 appear to have a long-term trend. It is worth to note the fact that the source was in the brightest state during the 2018 observation, both in the UV and X-rays. 
To quantify the variability exhibited by the source, we estimated the fractional variability amplitude $F_{var}$ \citep{Vaughan2003} in each energy band. The estimated count rates averaged over the exposure time and the $F_{var}$ values are quoted in Table~\ref{tab:table1}. The variability amplitudes of the X-ray variations are notably higher than those in the UV as is mostly the case for typical AGN. The UV variability are however still significant. 
%The table also reveal that the X-ray variability in XMM19 is higher on the average while the UV variability is lower compared to XMM13.

To investigate the X-ray/UV emission variability and their correlation, we employed three different widely-used cross-correlation analysis methods \citep[see e.g.,][]{Adegoke2019},
%where X-ray/UV correlation was searched for using $\sim100$ ks {\it XMM-Newton} data. And for UV, fast mode OM observation was used in the analysis. 
For XMM13, we considered the X-ray lightcurves of $\sim80\,\mathrm{ks}$ exposure (out of the $\sim 130\,\mathrm{ks}$ total exposure) that was simultaneous with the UVW1 exposures. In the top plot of Fig.~\ref{fig:fig1}, the vertical dotted line marks the duration of the X-ray/UV simultaneous observations for XMM13. For XMM18, we considered lightcurves spanning the entire duration in both the UV and the X-rays (shown in the middle plot of Fig.~\ref{fig:fig1}). Similarly, for XMM19 we used UV/X-ray lightcurves covering the entire duration of the outburst (shown in the bottom plot of Fig. \ref{fig:fig1}) for cross-correlation analysis.   

We first employed the Discrete Correlation Function (DCF) method of \citet{Edelson1988}. We used the {\tt python} implementation of the code {\tt pyDCF}\footnote{\url{https://github.com/astronomerdamo/pydcf}} which has been used widely on sampled or grouped data. In the runs, we imposed a lag bin width of $1.4\,\mathrm{ks}$. Determining the right lag bin with is always a compromise between statistical accuracy and resolution. A hard minimum for this value is typically set by the lightcurve time resolution.  
A careful look at Fig.~\ref{fig:fig2} indicates multiple DCF peaks in all cases. For XMM13 and XMM19, the peaks lie below $\pm{0.2}$ suggesting that there is no single peak that can be considered to be significant enough to indicate any time lag between the X-ray and the UV lightcurves. For XMM18 however, one of the DCF peaks extend up to $\sim0.4$, pointing to the possibility of a weak anti-correlation at $\sim30\,\mathrm{ks}$ ($\sim95\%$ confidence level). 
%For the three different X-ray bands, both XMM13 and XMM19 typically show similar behaviour, implying that there is no X-ray/UV correlation at any lag. 
XMM19 also tends to reveal a weak anti-correlation between the $2-10\,\mathrm{keV}$ X-ray and the UV bands around $25\,\mathrm{ks}$ at the 95\% confidence interval. Similarly weak X-ray/UV anti-correlation also appear to be present at $\sim15\,\mathrm{ks}$ in XMM13. 

In order to provide explicit support to the DCF analysis for the lack of correlation, we quantified the significance of any lag detection. For doing this, we simulated 10,000 pairs of lightcurves employing the widely used technique of {\tt bootstrap} \citep{1998PASP..110..660P}. From the discrete correlation functions of these {\tt bootstrap} samples, the 95\% and the 99\% confidence intervals were estimated by taking the $5.5^{th}/95.5^{th}$ and the $0.5^{th}/99.5^{th}$ percentiles respectively. The resulting plots are shown in Figure~\ref{fig:fig2}, where the dashed blue and red lines represent confidence intervals at the 95\% and 99\%, respectively. Figure~\ref{fig:fig2} clearly demonstrates that the DCF values against each time lag do not explicitly fall on the confidence intervals.

Further, we used another cross-correlation technique; the z - transformed discrete correlation function (ZDCF) \citep{Alexander1997,Alexander2013} to check for any UV/X-ray correlations and any associated time delays. ZDCF uses equal population binning and Fisher's z-transform. It is widely used to estimate the cross-correlation function of sparse, unevenly sampled lightcurves.  
In applying the ZDCF technique, we took the default minimum 11 data points per bin, and performed 1000 Markov Chain Monte Carlo (MCMC) simulations. As shown in Fig.~\ref{fig:fig3}, the overall pattern of the zdcf curves are generally similar to those from dcf, also showing multiple DCF peaks.

\begin{figure}
	\includegraphics[width=8cm, height=6cm]{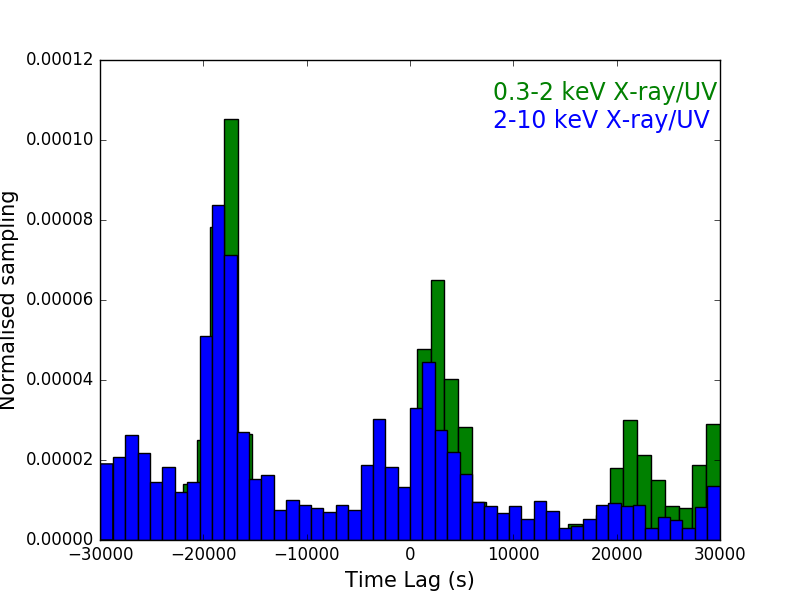}
	\vspace{0.0001cm}
    \includegraphics[width=8cm, height=6cm]{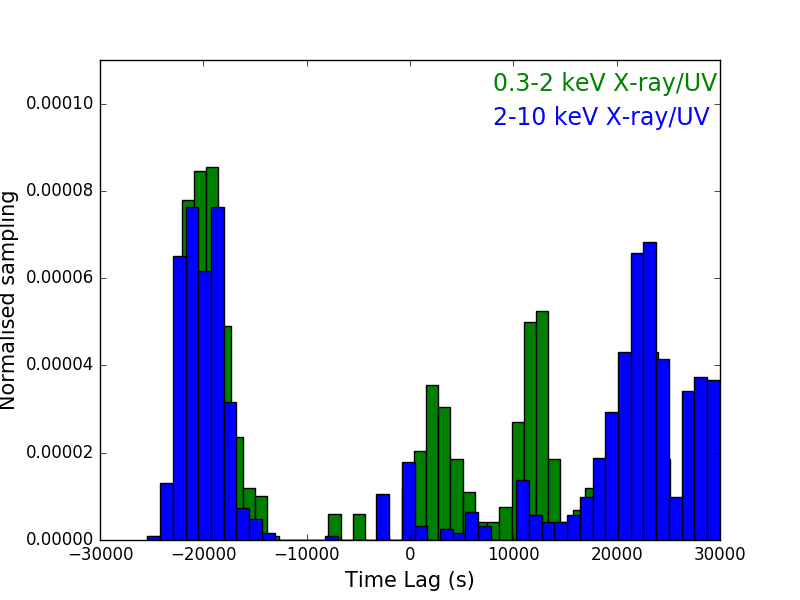}
    \vspace{0.0001cm}
	\includegraphics[width=8cm, height=6cm]{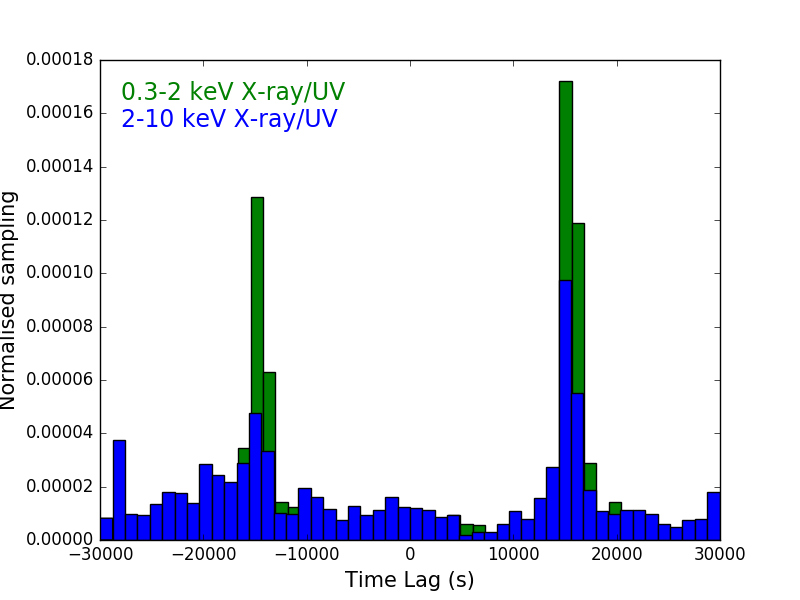}
	\caption{Results obtained from the JAVELIN code between X-ray and UV lightcurves. The top panel represents for XMM13,  middle panel represents for XMM18 and the bottom panel represents for XMM19 observation. The green and blue colors of the JAVELIN outcomes are presented for soft X-ray/UV and hard X-ray/UV bands, respectively.}
	\label{fig:fig5}
\end{figure}

Lastly, we used the JAVELIN code \citep{Javelin2011} to also check for correlated variability in the X-ray/UV lightcurves of XMM13, XMM18 and XMM19. The code uses a damped random walk process to model AGN continuum emission. It compares model lightcurves with observed ones to estimate the best-fitting lag. JAVELIN has been used to study correlated variability with various covariance functions \citep{javelin2013}. JAVELIN employs interpolation method and runs MCMC for large number of times, which derives posterior distribution of lags with shape in the form of Gaussian, where the peak indicates lag. This requires a simultaneous coverage of the lightcurves. We allowed 5000 MCMC runs and found that the Gaussian behaviour is totally absent in Mrk 1044. Instead, multiple peaks appears as obtained from the lightcurves at different time intervals (see Figure~\ref{fig:fig5}), which can not account for any significant correlation between X-ray and UV.

\subsection{Long time scale correlation}

\begin{figure}
	\includegraphics[width=8cm, height=6cm]{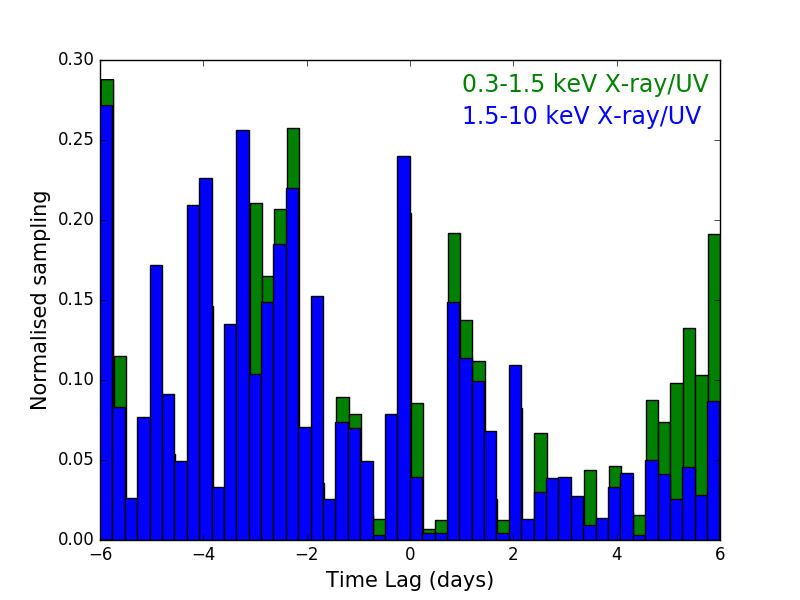}
	\caption{Representative {\it Swift} results produced by the application of JAVELIN to the UV and X-ray lightcurves. Respective plots for the soft and hard X-rays versus UV are shown by green and blue colors, respectively.}
	\label{fig:fig6}
\end{figure}

We use simultaneous \textit{Swift} XRT and UVOT observations of Mrk 1044 to search for any long-term correlations between the disc optical/UV and the coronal X-ray emissions. We used the techniques described above to search for any correlated variability between the lightcurves from the UVW1 filter against the three X-ray bands. For the DCF analysis, we considered the lag range of -7 to 7 days with a time bin-size of 1 day. The result obtained from the implementation of the DCF are shown in Fig~\ref{fig:fig4}. The DCF plot again shows multiple peaks with no significant lag between the UV and any of the X-ray bands.
%at the 99\% confidence level but tends to reveal a marginal correlation at the 95\% confidence level with a $\sim6-day$ delay and a marginal anti-correlation at $\sim5.3\,\mathrm{days}$, most prominent in the $0.3-10\,\mathrm{keV}$ X-ray/UV data. 
Also, as shown in Fig.~\ref{fig:fig4} (right plot), the ZDCF plot does not show any strong correlation between the UV and the X-ray bands. Application of the JAVELIN code (Fig.~\ref{fig:fig6}) also shows multiple peaks and no reliable correlations can be claimed for UV/X-ray variability. We went a step further to also check for any correlations between the other UVOT filters (UVM2, UVW2) against the combined $0.3-10\,\mathrm{keV}$ X-ray band but no significant correlations were detected.  

%{\bf Although the {\it XMM-Newton} exemplified the absence of any strong correlation between UV and X-rays on a short time scale, we undertook the monitoring observations taken with {\it Swift} to explore the correlation tests on longer time scale and to see the variability over each monitoring campaign. 

%The variability of the {\it Swift} based lightcurves was quantified similar to those of {\it XMM-Newton}. We estimated the fractional variability ($F_{var}$) for UV and three different X-ray bands 0.3--1.5 keV, 1.5--10 keV and 0.3--10 keV, which appears in Table~\ref{tab:table1}. We further noted that the X-ray variability is much higher than the UV. Also, the {\it Swift} UV shows higher variability over the {\it XMM-Newton} UV.

%%%
%%%%
%Correlation testes were performed for UV versus X-rays of three different energy bands by adopting the same techniques applied to the {\it XMM-Newton}. We furthermore took different UV bands (UVW1, UVM2, UVW2) for the tests. However, we present here results from UVW1 as their representative. The results obtained from the implementation of DCF are produced in Figure~\ref{fig:fig4}. We considered lag range of -7 to 7 days with time bin-size of 1 day. From a close visual inspection to the DCF product, it stand out that for all three X-ray bands versus UV, the CCF values lies under the 95\% and 99\% confidence intervals, and none of the plots produces any significant peak. Likewise, we do not see any strong correlation peak from the application of ZDCF (see Figure~\ref{fig:fig4}). Indeed, all the CCF values are within the error-bars. While applying JAVELIN, corresponding results to the other methods can be obtained as predicted. Output from the JAVELIN is presented in Figure~\ref{fig:fig6} which shows several peaks, yet, none of them is significant. Hence, conclusively, it can be said that consistent results produced by all three methods would be strong enough to provide evidence for the lack of X-ray/UV correlation.}

\section{X-ray Spectral analysis}
%\rm{Mrk~1044} has been shown to exhibit strong relativistic reflection features in its broadband spectrum including the broad component of the iron K$\alpha$ line. Using the $\sim130\,\mathrm{ks}$ long 2013 observation of the source (i.e. XMM13), \citet{Mallick2018} argued that relativistic disk reflection close to the central black hole is responsible for the prominent soft X-ray excess seen in the source. \textcolor{red}{In our analysis, we used the latest {\it XMM-Newton} observation obtained in 2019 to probe whether the spectra, in particular the soft X-ray excess, is also consistent with being reflection-dominated or an alternative explanation like thermal Comptonisation from a warm corona could equally be proffered. }
We used the latest {\it XMM-Newton} observation obtained in 2019 to probe whether the spectra, in particular the soft X-ray excess, is consistent with being reflection-dominated or an alternative explanation like thermal Comptonisation from a warm corona could equally be proffered.

The spectral analysis was carried out simultaneously for both the EPIC-pn spectra from XMM19 and FPMA/FPMB spectra from {\it NuSTAR} in the energy range of 0.3--50 keV. For the analysis we performed $\chi^{2}$ statistics using {\sc XSPEC} version 12.11.1 \citep{Arnaud} and estimated parameter errors at 90\% confidence interval. Neutral absorption along the line of sight was modelled with {\tt tbabs} using abundances from \citet{Wilms2000} and cross-sections from \citet{1996ApJ...465..487V}. 
%The line of sight column density $N_{H}$ was fixed at $2.88\times10^{20}\,\mathrm{cm^{-2}}$ \citep{2016A&A...594A.116H}.

%While the timing analysis plays an important role in searching correlated variability in different electromagnetic bands, we incorporate spectral analysis as well to investigate reflection dependency of the source in X-rays.   

%We attempted the spectral fitting procedure considering both 2019 {\it XMM-Newton} and 2018 {\it NuSTAR} observations together. The 0.3--10 keV EPIC-pn spectra from {\it XMM-Newton} and 3--50 keV FPMA/FPMB spectra from {\it NuSTAR} were fitted simultaneously in {\sc XSPEC} with its version 12.11.1 \citep{Arnaud}. We estimated the parameter errors at 90\% confidence level. % using {\sc XSPEC\_EMCEE}\footnote{\url{https://github.com/jeremysanders/xspec_emcee}} which is purely a python implementation of Markov Chain Monte Carlo (MCMC) ensemble sampler developed by Jeremy Sanders. Walkers of 50 and 10000 iterations were used, while burned the 1000 from the first. 

\subsection{Phenomenological model}
\begin{figure}
    \vspace{-0.3cm}
    \hspace{-0.5cm}
	\includegraphics[width=9cm, height=6.2cm]{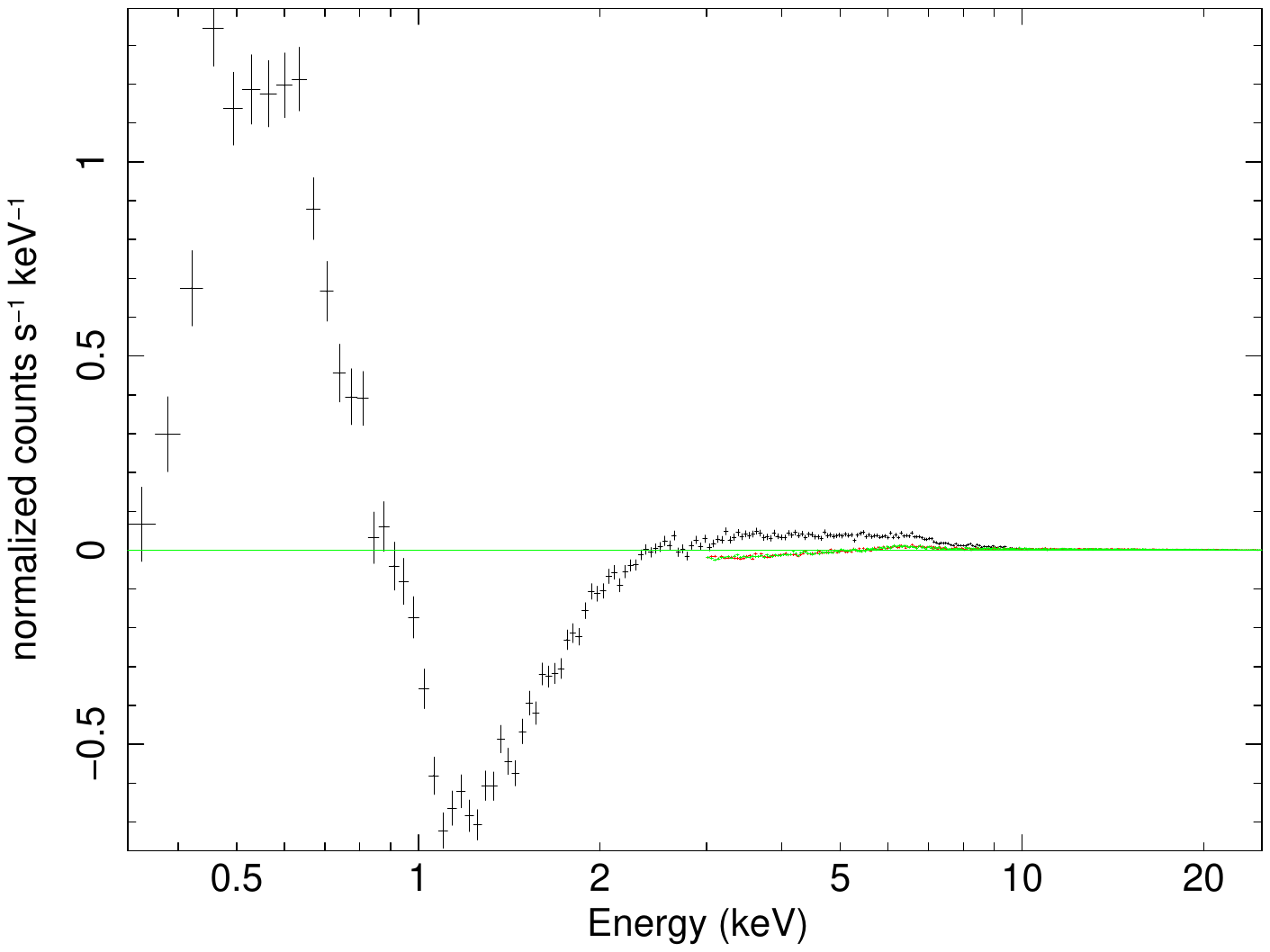}
 \caption{Residuals derived from a simple powerlaw fit to the  combined {\it XMM-Newton} and {\it NuSTAR} spectra. Appearance of the soft X-ray excess in the spectra is clear at the low energy. The spectra is shown up to 25 keV only for visual purpose.}
	\label{fig:fig7}
\end{figure}

\begin{figure}
    \vspace{-1.2cm}
    \hspace{-1.5cm}
	\includegraphics[width=13cm, height=8cm]{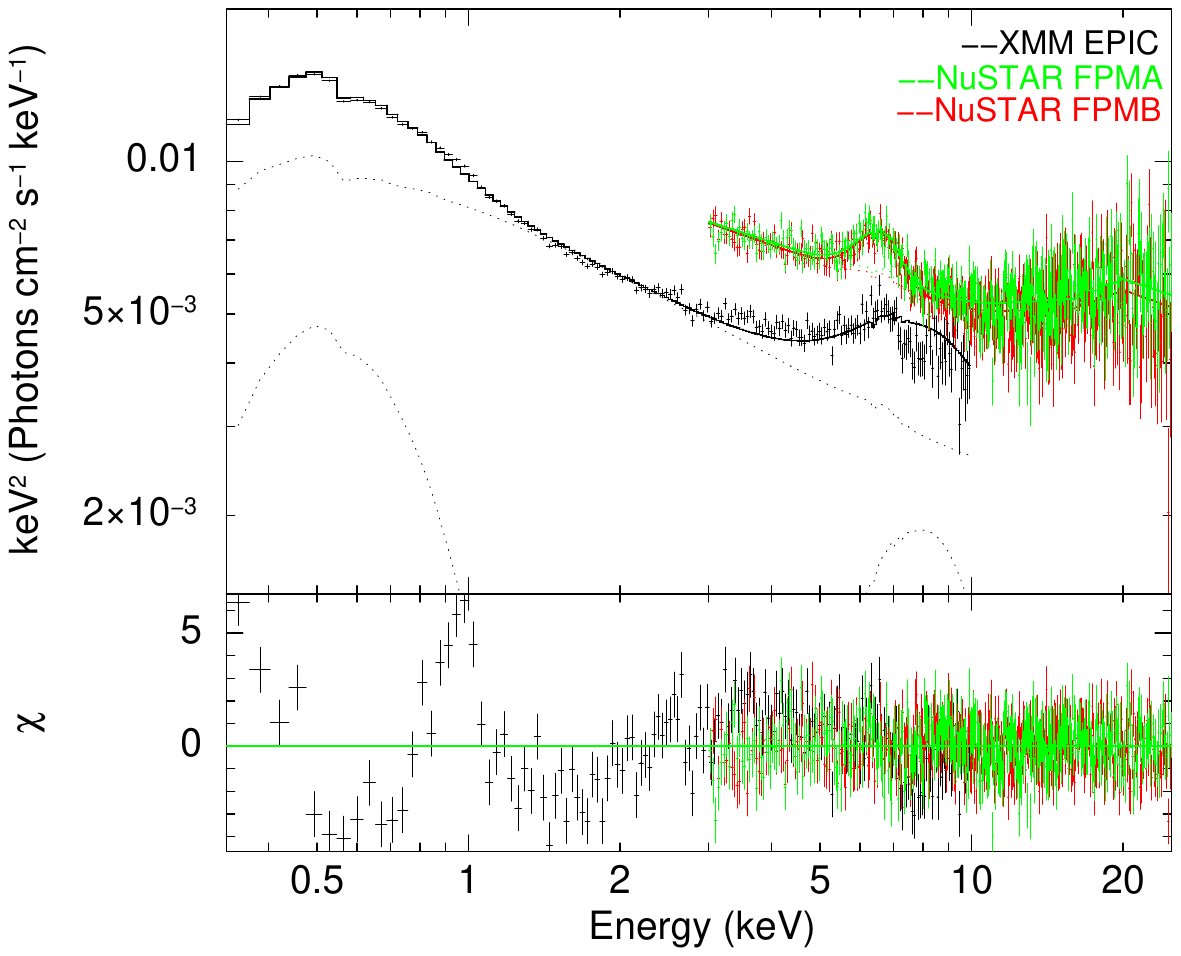}
 \caption{Spectra obtained from the fit with phenomenological model {\tt tbabs*zxipcf(bbody+powerlaw)}. The {\it XMM-Newton} and {\it NuSTAR} FPMA/FPMB spectra are displayed in black and green/red colors. For clarity in the data, the spectra is plotted up to 25 keV.}
	\label{fig:fig8}
\end{figure}

We started by fitting a simple absorbed {\tt powerlaw} model to the spectra in the $2-5\,\mathrm{keV}$ and the $7.5-10\,\mathrm{keV}$ ranges. This gives an acceptable fit with $\chi^{2}/dof = 109/83$. When extrapolated over the complete energy range, however, prominent features, including the soft X-ray below $\sim1\,\mathrm{keV}$, an absorption trough between $\sim1-2\,\mathrm{keV}$ and the broad component of the iron K$\alpha$ line, are evident (shown in Fig.~\ref{fig:fig7}). 
The soft excess was modelled by using a {\tt bbody} component which shows significant improvement in the fit. We then included the {\sc XSTAR}-based model {\tt zxipcf}, a model for partial absorption from partially ionised material, to fit for the absorption trough. Finally, a {\tt gauss} model was required to fit for the residual between $6-7\,\mathrm{keV}$. This provided a reasonable fit to the complete 0.3--50 keV spectra with $\chi^{2}/dof= 1496/973$. The fitted spectra is produced in Fig.~\ref{fig:fig8} which shows large residuals in the soft band, suggesting that the used model is not well favoured.

%%%%

%%%%
%%%%

%%%%
\begin{table*}
	\centering
	\caption{Spectral parameters from the combined fitting of the 0.3--10 keV {\it XMM-newton} and 3--50 keV {\it NuSTAR} FPMA/FPMB data using relativistic reflection model {\tt relxillCp}. Quoted errors are estimated at 90\% confidence level. Superscripts 1 and 2 represent that parameters are tied to those obtained from {\it XMM-Newton} and {\it NuSTAR} FPMA. Detailed description for the fitting can be found in section 4. }
	\label{tab:table2}
	\setlength{\tabcolsep}{15.0pt}
	\begin{tabular}{lcccccccc} % four columns, alignment for each
		\hline
		Model & Parameter & {\it XMM-Newton} &  {\it NuSTAR}   &{\it NuSTAR}    &  \\
                  &           & EPIC-pn          &      FPMA          & FPMB \\
		\hline
		Galactic absorption ({\tt tbabs}) & $N_H~(10^{20}~{\rm cm^{-1})}$ & $4.06_{-0.16}^{+0.16}$ & $4.06^1$& $4.06^1$\\
  
         \vspace{0.0cm}\\
        Partial covering ({\tt zxipcf})& $N_H~(10^{23}~{\rm cm^{-1})}$ & $47.38_{-38.62}^{+42.58}$ & $47.38^1$& $47.38^1$\\
                     \vspace{0.0cm}\\
                    & log$\xi$ & $2.73_{-1.33}^{+0.40}$ & $2.73^1$& $2.73^1$\\
                     \vspace{0.0cm}\\
                    & $C_{frac}$ & $0.21_{-0.04}^{+0.04}$ & $0.21^1$& $0.21^1$\\
	
		\vspace{0.0cm}\\
		Relativistic reflection ({\tt relxillCp})& $q_{1}$ & $> 9.70$  &  $ >9.60$ &  $9.60^2$ \\
		
        \vspace{0cm}\\
        & $R_{br}~(r_g)$ & $3.09_{-0.04}^{+0.05}$ & $3.09^1$ &  $3.09^1$\\
		
        \vspace{0cm}\\
        & $\Gamma$ & $2.41_{-0.01}^{+0.02}$ & $2.03_{-0.11}^{+0.08}$  & $2.00_{-0.05}^{+0.08}$\\
        
        \vspace{0cm}\\
        & $R$ & $5.20_{-0.53}^{+0.58}$ & $2.50_{-0.31}^{+0.76}$  & $2.50^2$\\

        \vspace{0cm}\\
        & $\theta$ & $34.00_{-1.33}^{+0.74}$ & $34^{1}$  & $34.00^1$\\

        \vspace{0cm}\\
        & $log\xi$ & $3.02_{-0.02}^{+0.02}$ & $3.02^1$  & $3.02^1$\\

        \vspace{0cm}\\
        & $kT_e$ & $100^f$ & $14.3_{-2.50}^{+3.30}$  & $14.30^2$\\

        \vspace{0cm}\\
        & $A_{fe}$ & $8.25_{-1.20}^{+1.04}$ & $8.25^1$  & $8.25^1$\\

        \vspace{0cm}\\
        & $\rm Norm~( 10^{-5})$ & $6.70_{-0.30}^{+0.27}$ & $3.52\pm1.75$  & $2.99\pm2.48$\\

        \vspace{0cm}\\
        & $\chi^2/{\rm dof}$ & $1260/1104$ & --  & --\\
	\hline
	\end{tabular}
 
        {\raggedright Here, unit of the ionisation parameter $\xi$ is $\rm erg~cm~s^{-1}$. $\rm C_{frac}$ represents the covering fraction. \par}   
\end{table*}
%%%%

\subsection{Physically motivated model}

%{\bf We applied physically-motivated models to the spectra to better understand the nature of the spectral features -- with particular focus on the soft X-ray excess -- exhibited by the source during this Epoch. We applied the thermal Comptonisation model and subsequently the relativistically blurred reflection model to understand the origin of the soft excess.

%We applied the thermal Comptonisation model {\tt optxagnf} \citep{Done2012} to test whether Comptonisation from a warm corona could equally provide an explanation to the origin of the soft excess. This model posit that the gravitation potential energy at each annulus of the disk is released as thermal blackbody emission up to the radius $R_{cor}$ of the corona. Below $R_{cor}$, this energy can no longer be efficiently thermalised, giving rise to an optically thick, warm plasma ($kT_{SE}\sim0.2\,\mathrm{keV}$) and an optically thin, hot corona ($kT_{e}\sim100\,\mathrm{keV}$). Thermal Comptonisation of disc photons in the warm corona is argued to be responsible for the soft X-ray excess. The complete model takes the form {\tt tbabs*zxipcf*(optxagnf+gauss)}, where {\tt zxipcf} and {\tt gauss} fit for the absorption feature and the broad iron K$\alpha$ line respectively. This model is unable to reproduce the soft excess, leaving prominent residuals in the soft band with $\chi^{2}/dof=4598/1115$.
We applied physically-motivated models to the spectra to better understand the nature of the spectral features -- with particular focus on the soft X-ray excess exhibited by the source. The relativistically blurred reflection model was applied to understand the origin of this feature and to describe the entire X-ray band.

%Next, we applied the relativistic reflection model. 
The widely used reflection model {\tt relxill} \citep{Dauser2014, Garcia2014} posits that features seen in the spectra of AGN, including the soft excess and the broad iron K$\alpha$ line arise from the relativistic blurring of ionised disc reflection due to the strong general relativistic effects close to the black hole. The model combines the capacity of the reflection code {\tt xillver} and the relativistic ray-tracing code {\tt relline}. For the disc material, the model assumes a powerlaw emissivity of the form $\epsilon\propto r^{-q}$, where $r$ and $q$ are the disc radius and the emissivity index respectively. 

\begin{figure}
    \vspace{-1.5cm}
    \hspace{-1.5cm}
	\includegraphics[width=13cm, height=8cm]{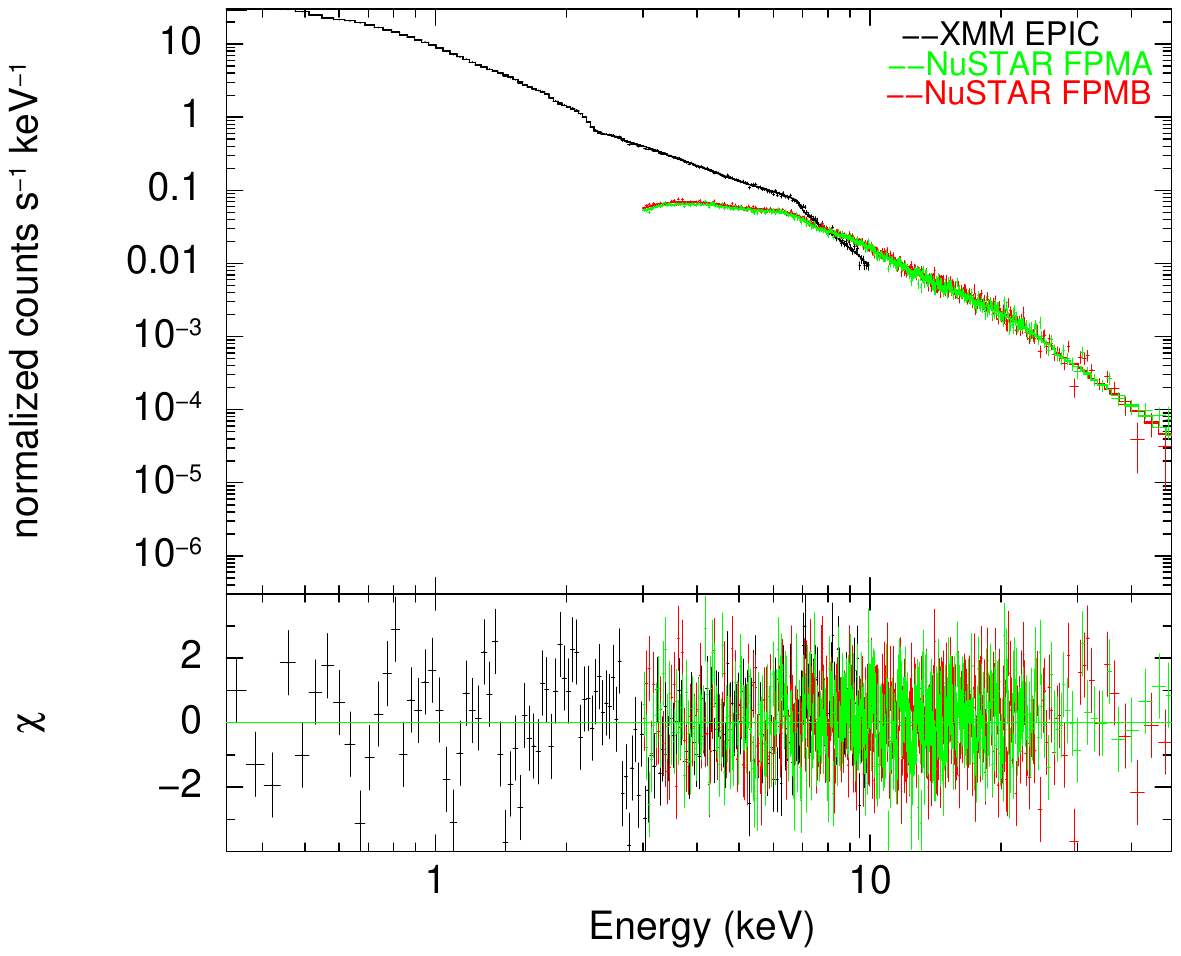}
 
	\vspace{-0.5cm}\caption{Joint 0.3--10 keV {\it XMM-Newton} and 3--50 keV {\it NuSTAR} spectra fitted with relativistic reflection model {\tt relxillCp}. Black color represents {\it XMM-newton} spectra while green and red colors represents {\it NuSTAR} FPMA and FPMB spectra, respectively. }
	\label{fig:fig9}
\end{figure}

Here, we used the {\tt relxillCp} flavour of the model to fit for the soft excess and the broad iron K$\alpha$ line and {\tt zxipcf} to model for the absorption feature between $\sim1-2\,\mathrm{keV}$. {\tt relxillCp} uses the primary continuum from the Comptonisation model {\tt nthcomp} \citep{1996MNRAS.283..193Z, 1999MNRAS.309..561Z} as opposed to the phenomenological {\tt powerlaw}. The complete model combination is {\tt tbabs*zxipcf*(relxillCp)}. During the fit, the inner and the outer radii were left frozen at their default values, while the break radius, inner emissivity index $q_{1}$ and the spin parameter a* were left free to vary. The spin takes the maximum value of 0.998 and a very steep emissivity is observed up-to break radius of $\sim 3r_g$. The fit parameters are quoted in Table~\ref{tab:table2} and the spectra in Fig.~\ref{fig:fig9}. The modeled reflection fraction for \textit{XMM-Newton} spectra high at $5.20_{-0.53}^{+0.58}$ compared to that from the \textit{NuSTAR} spectra which is $2.50_{-0.31}^{+0.76}$. This may not be unconnected to the fact that most of the reflection is contributed by photons at energies below $\sim2\,\mathrm{keV}$, making up the soft excess. The reflection modelling provided a better fit to the data with $\chi^2$/dof = 1260/1104 and improved residuals than the phenomenological one, implying that reflection is more acceptable than the Comptonisation. Contemporaneous fit by the warm Comptonisation and reflection models is also reported in the recent work of \citet{Yu2023}, where six NLS1 AGNs are analysed. As the authors suggest, spectrum of a single source can be better described by the relativistic reflection, although the models give very similar fit-statistic, making difficult to determine which scenario should be more favourable for the origin of the soft excess.

\section{Discussion}
\begin{figure*}
\centering
        \hspace{-0.5cm}
		\includegraphics[width=5cm, height=5cm]{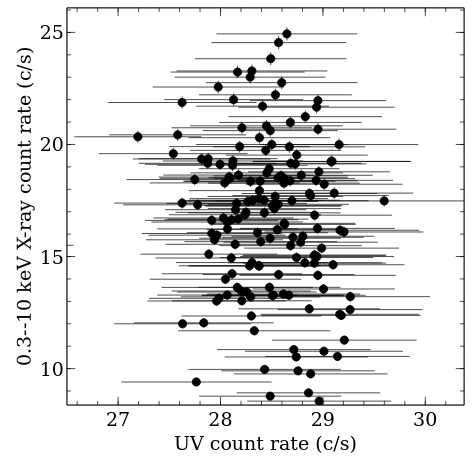}
	      \hspace{0.1cm}
        \includegraphics[width=5cm, height=5cm]{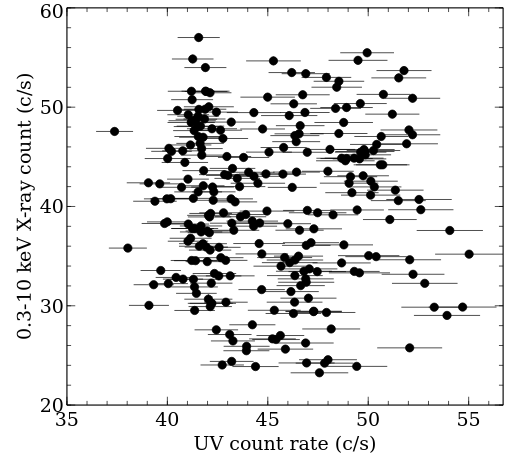}
		\includegraphics[width=5cm, height=5cm]{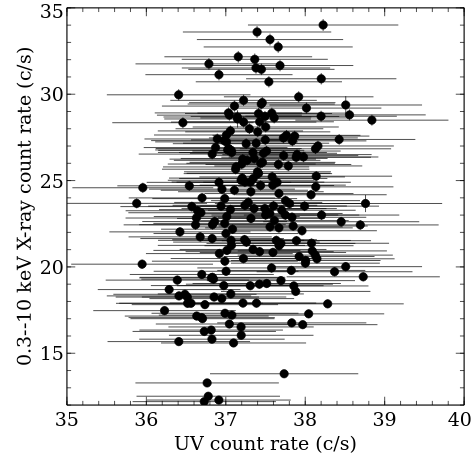}
		\caption{Plots for UV flux versus X-ray flux constructed from  XMM13 (left), XMM18 (middle) and XMM19 (right) lightcurves. It is apparent that no correlation exists between X-ray and UV emissions.
		}
		\label{fig:count_count.png}
\end{figure*}

To investigate X-ray/UV correlated variability in {Mrk 1044}, we carried out cross-correlation analysis using three long {\it XMM-Newton} and numerous {\it Swift} monitoring observations with simultaneous UV and X-ray data. Our analysis revealed that there is no significant correlation between emission in the UV and the X-ray energy bands on short as well as long time scales. Although there appears to be anti-correlation features in the CCFs, they are considered insignificant. Plots from the JAVELIN code (Fig. \ref{fig:fig5}) show multiple correlation peaks which are also insignificant.
In order to cross-check the results produced by each correlation technique and thereby to further validate that X-ray/UV correlation is nonexistent, we constructed plots for the X-ray count rates against the UV from the {\it XMM-Newton} lightcurves (see Fig.~\ref{fig:count_count.png}). Clearly, no trend appears between X-ray flux and UV flux. 

Besides timing analysis, we conducted a broadband spectral analysis to search for X-ray reflection dependency of the source. Report appeared in \citet{Mallick2018} revealed the presence of strong soft excess based on reflection spectroscopy of the 0.3--50 keV spectra given by the 2013 {\it XMM-Newton} and 2016 {\it NuSTAR} observations. Further, our detailed analysis for the {\it XMM-Newton} along with one of the longest {\it NuSTAR} observation shows that including the soft excess the broadband spectra can be described by the relativistic reflections

%However, the more recent, nearly simultaneous and long observations are anticipated in placing strong ground on the reflection dependency.}
%%
%{\bf From the $\sim$100 ks {\it XMM-Newton} observations of Mrk 493 \citet{Adegoke2019} reported X-ray/UV correlation consistent with a 5 ks lag, where the UV variability was found to lead the X-ray variability. The correlation was confirmed having obtained consistent results from DCF, ZDCF and JAVELIN, respectively. This X-ray lag has been interpreted to be due to Comptonisation -- scattering of disk photons into X-rays.}
%%	
%{Observations of roughly same duration ($\sim$130 ks) and similar techniques were used in this study for Mrk 1044. The DCF does not reveal correlation at any time lag at low significance (95\% confidence), while it is more clear at higher significance (99\% confidence). By applying ZDCF in which the correlation function is estimated with significant improvement over DCF, the lack of correlation can be well noticed. At last, results obtained from the use of JAVELIN code reinforce the DCF and ZDCF findings. The observed lack of correlation is not usual. However, such non-detections were observed in reflection dominated NLS1s 1H 0707-495 and IRAS 13224-3809 by \citetalias{Robertson2015} and \citetalias{Buisson2018}, respectively.  }

Sources with evidence of strong reflection in their X-ray spectra should normally be expected to exhibit X-ray/UV CCF delay consistent with reprocessing -- where emission in the UV lag behind those in the X-rays. This is because the same X-ray powerlaw emission illuminating the disc is not only responsible for the observed reflection features, but should also be responsible for the reprocessed lower energy UV photons in the accretion disc \citep{Edelson2015, Pal2017, Buisson2017, Lobban2018, Kammoun2019}. Most AGNs are expected to have a standard accretion disc \citep{Czerny2018} and we assume that this is also true for Mrk 1044 which has an Eddington ratio of $\sim 0.3$. Thus, following \citet{Adegoke2019}, we can estimate the radius at which the UV radiation is emitted using the relation $T \propto R^{-3/4}$. 
In the reprocessing scenario, the light travel time -- corresponding to the expected time delay -- between the central illuminating X-rays and the disc UVW1 photons observed at $\lambda_{eff}=2910\textup{\AA}$ should follow the relation $t_{lc} \propto \lambda_{eff}^{4/3}$ \citep{McHardy2016} and can be expressed as 
\begin{equation}
t_{lc} \approx 2.6 \times 10^5 \left(  \frac{\lambda_{eff}}{3000\mathring{A}} \right)^{4/3} \left( \frac{\dot{M}}{{\dot{M}_{Edd}}}\right)^{1/3}  
\left( \frac{{M_{BH}}}{10^8M_\odot}\right)^{2/3}              
%x=\frac{-b\pm\sqrt{b^2-4ac}}{2a}.
\label{eq:eq1}
\end{equation}\\

We analysed the set of all three {\it XMM-Newton} observations to measure the accretion rate. The accretion rate (${\dot{m}}$) was calculated from the ratio of bolometric luminosity ($L_{bol}$) to the Eddington luminosity ($L_{Edd}$), where we obtained the $L_{bol}$ by applying bolometric correction to the 2--10 keV luminosity following the prescription outlined in \citet{Duras2020}. XMM13 yields the accretion rate at $\sim 0.25$ whereas $L_{bol} = 10.08 \times 10^{43}$ erg/s and 2-10 keV luminosity $L_{\rm 2-10 ~ keV} = 7.85 \times 10^{42}$ erg/s. This becomes $\sim 0.43$ in XMM18, where $L_{bol} = 16.52 \times 10^{43}$ erg/s and $L_{\rm 2-10 ~ keV} = 12.82 \times 10^{42}$ erg/s. Further, accretion rate of $\sim 0.31$ is obtained from XMM19, with $L_{bol}= 12.06 \times 10^{43}$ erg/s and $L_{\rm 2-10 ~ keV} = 9.37 \times 10^{42}$ erg/s. Inserting these accretion rate values (${\dot{m}}$= 0.25, 0.31 and 0.43) in equation~\ref{eq:eq1},  $t_{lc}$ can be obtained to be at $\sim15.05$ ks, $\sim16.17$ ks and $\sim18.03$ ks, respectively.
%{\bf Considering a near Eddington-scaled accretion rate of 0.3 \citep{Laha2018}, this gives $t_{lc}\sim16\,\mathrm{ks}$ and for an accretion rate of the order of high Eddington (i.e. ${\dot{m}}=16.0$; \citet{Du2015}), $t_{lc}\sim60.7\,\mathrm{ks}$.} This is assuming the reprocessing time on the disc is negligible.

Further, we derived the soft X-ray strengths from the ratios of 0.3--1 keV luminosity of black body component to the Eddington luminosity ($ SX1=L_{\rm bb~(0.3-1~ keV)}/L_{Edd}$) and 0.5--2 keV black body flux to the continuum flux ($ SX2=(F_{bb}/F_{con})_{\rm 0.5-2~keV}$). While the strength $SX1$ takes the values at $0.01 \pm 0.007$, $0.02 \pm 0.015$, and $0.03 \pm 0.01$, $SX2$ at $1.10 \pm 0.02$, $1.50 \pm 0.01$ and $1.48 \pm 0.006$ against the accretion rates of $\sim0.25$, $\sim0.31$ and $\sim0.43$ corresponding to the 2--10 keV luminosity $L_{\rm 2-10 ~ keV}$ of $7.85 \times 10^{42}$ erg/s, $9.37 \times 10^{42}$ erg/s and $12.82 \times 10^{42}$ erg/s, respectively. This, however, indicates that the soft X-ray strength does not vary significantly, and we do not see any strong correlation or anti-correlation with ${\dot{m}}$ and $L_{\rm 2-10 ~ keV}$. It may be pointed up that more number of observations spanning a wide range of luminosity would be required if one performs a robust analysis to search for any correlation and/or anti-correlation.

%%%%
If the observed UV emission is due to reprocessing of X-rays in the disc, then the UV lightcurves should show variability associated with the X-ray lightcurves. In the case where reprocessing of X-rays in the disc is significant, such a disc will be hotter than when heated solely by accretion. Supposing that \rm{Mrk~1044} was illuminated by a compact X-ray source located above the disc at a height of $\sim20-200\,R_{g}$, then additional UVW1 flux ($\sim30-40\%$) will be generated \citep[e.g.,][]{Robertson2015}. The observed UVW1 variability measured for \rm{Mrk~1044} is $\sim$2\% from the {\it XMM-Newton} and $\sim17$ \% from the {\it Swift} which are much less than the variability in X-rays, implying that the disc in this source may be illuminated anisotropically by the X-ray source consequently giving rise to the lack of correlation in the variability seen in both bands. Variability in the geometry and structure of the corona may also introduce additional complexity to the X-ray/UV lightcurve variability \citep[e.g.,][]{2015MNRAS.449..129W}
%%
%{\bf The observed weak anti-correlation measured in the source is most likely arising due to the strong general relativistic effects in the immediate vicinity of the black hole. This can, however, be noted from the reflection dependency of soft X-ray excess as shown by the spectroscopic analysis.} 
Depending on how high the compact X-ray source is above the disc, gravitational light-bending would make the amount of radiation from the isotropic emitting source as seen by the disc to look different than that seen by a distant observer \citep[e.g.,][]{Miniutti2004}. Alternatively, if the X-ray source were the base of a jet moving at a significant fraction of the speed of light, then beaming effects would make emission from the isotropic X-ray source as seen by the disc appear to be anisotropic \citep[e.g.,][]{1997MNRAS.290L...1R}

%Our analysis of the \textit{XMM-Newton} data of \rm{Mrk~1044} revealed no evidence for correlation between its UV and X-ray emitting regions. 
Timing analysis of a number of other Seyfert 1 AGN have showed no X-ray/UV correlations on a similar timescale as considered for \rm{Mrk~1044}. A few of them exhibit similar spectral properties to \rm{Mrk~1044}. One such source is \rm{1H~0707-495} having a mass of $2\times10^{6}\,M_{\odot}$, also known to possess a reflection-dominated X-ray spectra. For this AGN, \citetalias{Robertson2015} argued that strong GR effects and significant light-bending in the innermost regions of the accretion disc could prevent substantial X-ray photons from reaching far out into the disc leading to the non-detection of X-ray/UV correlation. Another X-ray bright NLS1 AGN IRAS 13224-3809 has similar spectral properties to Mrk 1044, showing that X-ray is highly variable than UV \citepalias{Buisson2018}. As the authors suggest, any possible X-ray/UV correlation may be blocked from detection due to higher X-ray variability.

It is possible that scattering and/or absorption between the corona and the disc or changes in the coronal geometry brings about a difference between the X-ray variability viewed by the disc and that along our line of sight. The underlined change in the coronal geometry is typical when it moves closer to the black hole as seen in the reflection dominated NLS1 galaxy Mkn 478 \citep{Barua2022}, or due to the corona collapsing down onto the black hole. This is akin to the coronal emissions in a larger fraction to focus on the innermost disc, while only a small fraction is to be received by the outer disc, probably causing the enhanced complexity in the X-ray variability and thereby leading to the non-detection of the correlation. 
Future X-ray instruments with larger collecting area may help to map the corona and would allow a better measurement of the X-ray irradiation of the disc in addition to the X-ray flux along our line of sight.

\section{conclusion}
Our analysis for NLS1 AGN Mrk 1044 based on three long {\it XMM-Newton} observations does not reveal any significant correlation on short timescales between the X-ray and UV variability in their lightcurves. Similarly, the {\it Swift} monitoring observations reveal that correlation lacks on longer timescale. This non-detection is not usual, but not even unpredictable as consistent findings have been reported for NLS1 AGNs 1H 0707-495 and IRAS 13224-3809 which show similar spectral properties to Mrk 1044. One plausible cause for this lack of correlation is attributed to the light bending effect -- mostly seen in reflection dominated sources. Else, expected to be due to higher X-ray variability over UV, blocking the detection of the correlation. 
\begin{acknowledgments}
We thank the anonymous referee for the constructive comments and suggestion which have improved the manuscript. We thank Prof. Gulab C Dewangan for his useful inputs in the manuscript. SB acknowledge Inter-University Centre for Astronomy and Astrophysics (IUCAA) for the support. VJ acknowledges the support provided by the  Department of Science and Technology under the ``Fund for Improvement of S \& T Infrastructure (FIST)" program (SR/FST/PS-I/2022/208). This research has made use of data obtained through the High Energy Astrophysics Science Archive Research Center Online Service, provided by the NASA/Goddard Space Flight Center.
\end{acknowledgments}

\vspace{5mm}
\facilities{ XMM-Newton}
\software{pyDCF \citep{Edelson1988},  ZDCF \citep{Alexander1997}, JAVELIN \citep{Javelin2011}}

 %\url{http://journals.aas.org/authors/aastex.html}

%\appendix
%\section{Appendix information}
\bibliography{reference}{}
\bibliographystyle{aasjournal}
\end{document}